\documentclass[aps,prb,twocolumn,superscriptaddress,bibliography]{revtex4-2}
\usepackage{amsfonts}
\usepackage{mathrsfs}
\usepackage{amsmath}
\usepackage{color}
\usepackage{xcolor}
\usepackage{graphicx}
\usepackage{bm}
\usepackage{amssymb}
\usepackage{xspace}
\usepackage{epstopdf}
\usepackage{dcolumn}
\usepackage{longtable}
\usepackage{multirow}
\usepackage{float}
\usepackage{comment}
\usepackage{booktabs}

\usepackage[colorlinks=true, letterpaper=true, pdfstartview=FitV,  linkcolor=blue, citecolor=blue, urlcolor=blue]{hyperref}

\makeatother

\begin{document}

\title{Berry curvature dipole and nonlinear Hall effect in two-dimensional Nb$_{2n+1}$Si$_n$Te$_{4n+2}$}

\author{Yiwei Zhao}
\affiliation{State Key Laboratory for Mechanical Behavior of Materials, Xi'an Jiaotong University, Shaanxi 710049, China}
\affiliation{Research Laboratory for Quantum Materials, Singapore University of Technology and Design, Singapore 487372, Singapore}

\author{Jin Cao}
\email{caojin.phy@gmail.com}
\affiliation{Centre for Quantum Physics, Key Laboratory of Advanced Optoelectronic
	Quantum Architecture and Measurement (MOE), Beijing
	Institute of Technology, Beijing 100081, China}
\affiliation{Research Laboratory for Quantum Materials, Singapore University of Technology and Design, Singapore 487372, Singapore}

\author{Zeying Zhang}
\email{zzy@mail.buct.edu.cn}
\affiliation{College of Mathematics and Physics, Beijing University of Chemical Technology, Beijing 100029, China}
\affiliation{Research Laboratory for Quantum Materials, Singapore University of Technology and Design, Singapore 487372, Singapore}

\author{Si Li}
\affiliation{School of Physics, Northwest University, Shaanxi 710127, China}

\author{Yan Li}
\affiliation{State Key Laboratory for Mechanical Behavior of Materials, Xi'an Jiaotong University, Shaanxi 710049, China}

\author{Fei Ma}
\email{mafei@mail.xjtu.edu.cn}
\affiliation{State Key Laboratory for Mechanical Behavior of Materials, Xi'an Jiaotong University, Shaanxi 710049, China}

\author{Shengyuan A. Yang}
%\email{shengyuan\_yang@sutd.edu.sg}
\affiliation{Research Laboratory for Quantum Materials, Singapore University of Technology and Design, Singapore 487372, Singapore}

%\date{\today}

\begin{abstract}
Recent experiments have demonstrated interesting physics in a family of two-dimensional (2D) composition-tunable materials Nb$_{2n+1}$Si$_n$Te$_{4n+2}$. Here, we show that owing to its intrinsic low symmetry, metallic nature, tunable composition, and ambient stability, these materials offer a good platform for studying Berry curvature dipole (BCD) and nonlinear Hall effect. Using first-principles calculations, we find that BCD exhibits pronounced peaks in 
monolayer Nb$_{3}$SiTe$_{6}$ ($n=1$ case). Its magnitude decreases  monotonically  with $n$ and completely vanishes in the $n\rightarrow\infty$ limit. This variation manifests a special hidden dimensional crossover of the low-energy electronic states in this system. The resulting nonlinear Hall response from BCD in these materials is discussed.
Our work reveals pronounced geometric quantities and nonlinear transport physics in Nb$_{2n+1}$Si$_n$Te$_{4n+2}$ family materials, which should be readily detected in experiment.
\end{abstract}
\maketitle

\section{Introduction}

The Hall effects, in which a transverse current $j_\text{H}$ is induced by a longitudinal driving $E$ field, are of fundamental importance in condensed matter physics~\citep{von1986quantized,nagaosa2010anomalous,sinova2015spin}. At linear order, i.e., with $j_\text{H}\sim E$, the Hall effect requires the broken time reversal symmetry $\mathcal{T}$, which can be achieved either by an applied magnetic field or by intrinsic magnetism. This constraint is loosened when considering Hall responses at nonlinear order, as the non-equilibrium electron distribution driven by $E$ field already breaks $\mathcal{T}$ at its first order. Focusing on the second-order response, in nonmagnetic materials and in the absence of magnetic field, Sodemann and Fu proposed a Berry curvature dipole (BCD) contribution to the nonlinear Hall current $j_\text{H}\sim E^2$ within the semiclassical theory framework~\citep{sodemann2015quantum}. Their work attracted great interest in the past few years, and the effect has been successfully detected in several material systems~\citep{ma2019observation,kang2019nonlinear,Son2019,Battilomo2019,Kim2019,Dzsaber2021,He2021_Bi2Se3,qin2021strain,Huang2022,Zhang2022_ferro,Sinha2022,Du2018_bilayerWTe2,Zhang2018_Weyl,you2018berry,zhang2018electrically,Facio2018,xiao2020two,Zeng2021_Weyl,Wawrzik2021,Lu2021_review}.  It was suggested that this effect offers a new mechanism for nonlinear applications, such as frequency-doubling and rectification~\citep{pacchioni2019hall,zhang2021terahertz,kumar2021room}.

For experimental study, two-dimensional (2D) materials have advantages in their great tunability. For example, the Fermi level in 2D materials can be readily tuned via gating technique to a large extent not possible in 3D bulk materials~\citep{chen2010gate,ma2019observation}.
However, regarding BCD and its induced nonlinear Hall effect, the constraint from crystalline symmetry in 2D is rather stringent. It was shown that the largest symmetry in 2D that allows for a nonzero BCD is a single {in-plane} mirror line~\citep{sodemann2015quantum}. Hence, to realize the effect, one has to choose crystals with very low symmetry, which are rather limited, or takes extra effort to exert strain or twist on the crystal to lower the symmetry. This severely hinders the experimental study.

Recently, the family of composition-tunable materials Nb$_{2n+1}$Si$_n$Te$_{4n+2}$ have attracted interest from both theory and experiment~\citep{Li2018_NbSiTe,Sato2018_NbSiTe,Yang2019_NbSiTe,Zhu2020_NbSiTe,Wang2021_NbSiTe,Zhang2022_NbSiTe}. In the bulk form, these materials are van der Waals layered materials. Their high-quality 2D layers can be obtained by mechanical exfoliation method~\citep{Hu2015}. The special feature of this family is the tunable composition embodied by the integer $n$~\citep{li1992synthesis,monconduit1993synthesis,evain1994modulated,van1994superspace}. For each $n$, the system is a stoichiometric crystal, and the physical properties have an interesting dependence on $n$. For example, it was shown that in a 2D monolayer, for finite $n$, the material is nonsymmorphic nodal-line semimetal~\citep{Li2018_NbSiTe}; whereas the $n\rightarrow \infty$ limit, i.e., the material Nb$_2$SiTe$_4$, is a narrow-gap semiconductor~\citep{zhao2019nb2site4,Wang2021_NbSiTe}. With increasing $n$, the low-energy states at Fermi level exhibits a dimensional change from 2D-like states to 1D-like states~\citep{Zhu2020_NbSiTe}.

We note that 2D Nb$_{2n+1}$Si$_n$Te$_{4n+2}$ materials actually offer a good platform to explore BCD related physics. First, except for the $n\rightarrow \infty$ limit, all members of the family have a sufficiently low symmetry to allow an intrinsic BCD, without the need of applied strain. Second, they offer an opportunity of systematic investigation of the evolution of BCD with the tunable composition. Third, these 2D materials are stable at ambient conditions~\citep{Zhu2020_NbSiTe}, which facilitates experimental study as well as possible applications.

Motivated by the above considerations, in this work, we theoretically study BCD and nonlinear Hall effect in monolayer Nb$_{2n+1}$Si$_n$Te$_{4n+2}$ materials. With first-principles calculations, we show that the $n=1$ case, i.e., Nb$_{3}$SiTe$_{6}$, possesses a pronounced BCD. The magnitude can reach {1.54 $\rm\AA$} in the hole doped case, larger than previously reported values in {2D $T_d$-WTe$_2$~\citep{you2018berry}, strained NbS$_2$~\citep{xiao2020two} and WSe$_2$~\citep{you2018berry}. With increasing $n$, the BCD peaks in the spectrum show a monotonic decrease and eventually vanish in the $n\rightarrow\infty$ limit. This behavior can be understood from two perspectives. One is from the symmetry perspective, and the other is from the dimensional evolution of the electronic states. The latter view manifests that although structurally, these materials are strongly bonded in both directions in 2D, electronically, the states exhibit a dimensional crossover from 2D to 1D. This hidden crossover dictates the change in BCD. The key features of the results are further captured by our
constructed tight-binding models for this family of materials. To guide experiment, we discuss properties of the nonlinear Hall response arising from BCD. Our work reveals interesting properties of Nb$_{2n+1}$Si$_n$Te$_{4n+2}$ family materials and suggests them as a suitable platform to explore BCD and nonlinear Hall physics.

\section{Computation method}
Our first-principle calculations were based on the density functional theory (DFT), performed by using the Vienna \emph{ab initio} simulation package~\citep{VASP1,VASP2,VASP3}. The ionic potentials were treated by using the projector augmented wave method~\citep{PAW}. The exchange-correlation functional was treated by the generalized gradient approximation~\citep{GGA} in the scheme by Perdew, Burke, and Ernzerhof~\citep{PBE}. The plane-wave cutoff energy was set to be 400~eV, and a $10\times4\times1$ $\Gamma$-centered $k$-point mesh was used for the Brillouin zone (BZ) sampling. The convergence criteria for the total energy and the force were set to be $10^{-6}$~eV and 0.01~eV/$\rm\AA$, respectively. To avoid artificial interaction between periodic images, a vacuum space of 15~$\rm\AA$ thickness was added. Spin-orbital coupling (SOC) was included in all calculations. Based on the band structure calculation, an \emph{ab initio} tight-binding model was constructed by using the Wannier90 package~\citep{WANNIER90_CPC}. The $d$ orbitals of Nb atoms and $p$ orbitals of Te atoms were used as the initial guess of the local basis. The  BCD was calculated based on this \emph{ab initio} tight-binding model. In evaluating BCD, we set $T=100$~K in the Fermi distribution function.

\begin{figure}
\begin{centering}
\includegraphics[width=8.6cm]{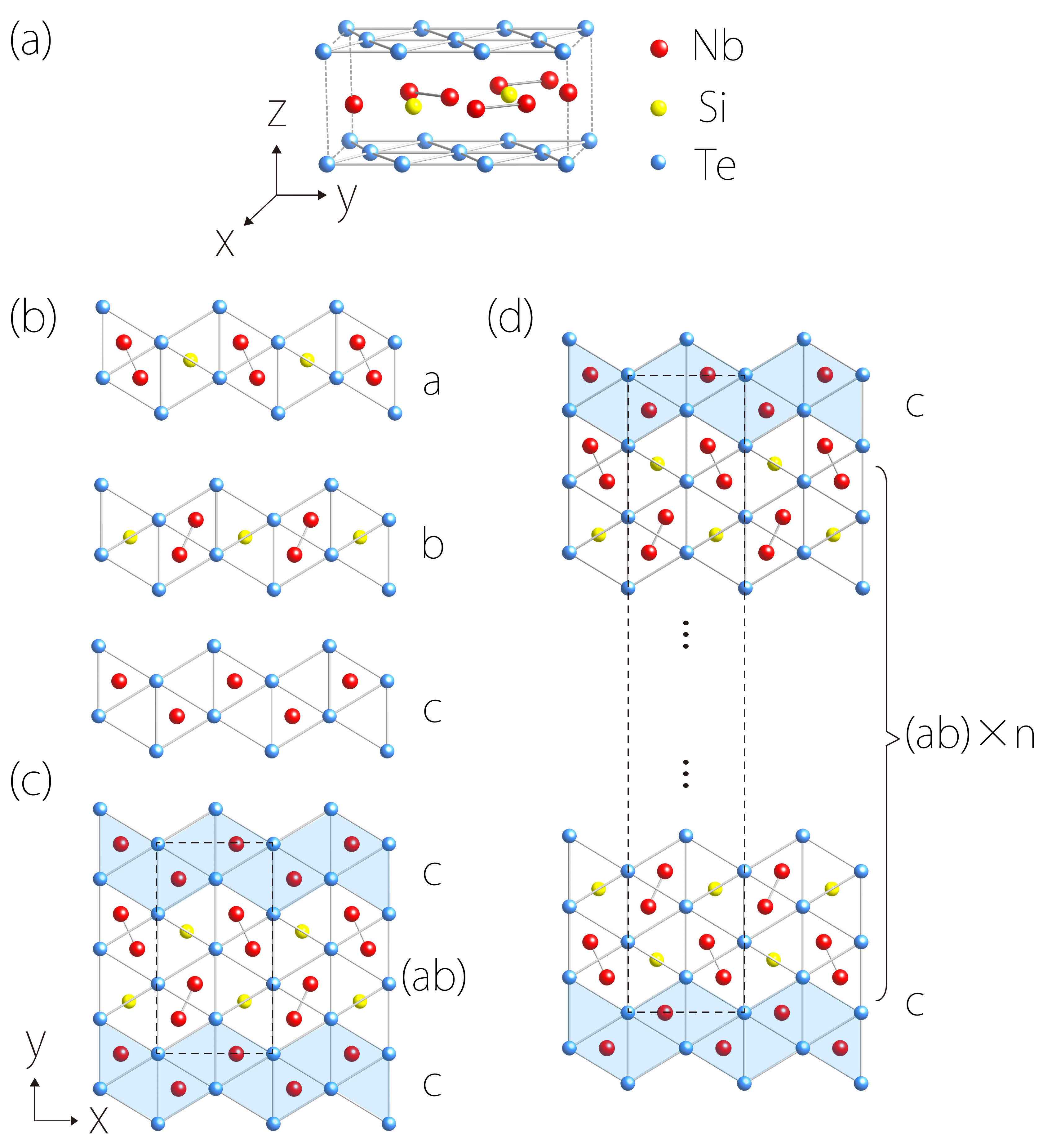}
\par\end{centering}
\caption{\label{Fig_1}{(a) Lattice structure of monolayer Nb$_3$SiTe$_{6}$. (b) The three building blocks of of Nb$_{2n+1}$Si$_n$Te$_{4n+2}$ family materials: $a$, $b$ and $c$ chains. (c) Top view of $n=1$ case (Nb$_3$SiTe$_{6}$). The dashed box marks the unit cell. (d) Nb$_{2n+1}$Si$_n$Te$_{4n+2}$ can be constructed by $n$ copies of $(ab)$ chains and one $c$ chain in a unit cell (the dashed box). }}
\end{figure}

\section{Crystal and electronic structures}

The Nb$_{2n+1}$Si$_n$Te$_{4n+2}$ family materials were first synthesized in the 1990s by {chemical vapour transport method}~\citep{li1992synthesis}. The lattice structures of their 2D monolayers are illustrated in Fig.~\ref{Fig_1}. Here, each monolayer consists of three atomic layers: the middle layer containing Nb and Si atoms is sandwiched by two Te layers [Fig.~\ref{Fig_1}(a)]. From the top view [see Figs.~\ref{Fig_1}(b)-\ref{Fig_1}(d)}], these materials can be viewed as composed of three building blocks, which are conventionally called the $a$, $b$, $c$ chains. As shown in Fig.~\ref{Fig_1}(b), $a$ and $b$ chains contain Si atoms and share the same composition of NbSi$_{1/2}$Te$_2$, whereas the $c$ chain does not contain Si and has the composition of NbTe$_2$. Assuming these chains are along the $x$ direction [as in Fig.~\ref{Fig_1}(c)], then $a$ and $b$ are connected by a glide mirror operation $\tilde{M}_y=\{M_y|\frac{1}{2}0\}$, and in these materials they always appear together.
Members of this family are formed by assembling these chains along the lateral direction ($y$) in a periodic manner, such that Nb$_{2n+1}$Si$_n$Te$_{4n+2}$ corresponds to the arrangement of $(ab)_nc$. Namely, in a period, we have one $c$ chain and $n$ copies of $(ab)$ chains, as illustrated in Fig.~\ref{Fig_1}(d). In the $n=\infty$ limit, there is no $c$ chain in the structure any more, and we reach the composition of Nb$_{2}$SiTe$_{4}$.

Our optimized lattice parameters for $n=1,2,3,\infty$ are listed in Table~\ref{tab:1}. These values are in good agreement with experiment and previous calculations~\citep{Li2018_NbSiTe,Sato2018_NbSiTe,Zhu2020_NbSiTe}. We also note that for members with finite $n$, they all have the space group symmetry {$Pmc2_1$}, with {$C_{2v}$} point group. In comparison, Nb$_{2}$SiTe$_{4}$ with $n=\infty$ has a larger space group {$Pbam$} and a point group {$D_{2h}$}. The main difference is the extra glide mirror  $\tilde{M}_x=\{M_x|0\frac{1}{2}\}$ for $n=\infty$ case but not for any finite $n$. From Fig.~\ref{Fig_1}(c), one can see that it is the $c$ chains that break the $\tilde{M}_x$ symmetry which holds for $(ab)$ chains.

\begin{table}
\centering
\caption{{Optimized lattice parameters and the corresponding symmetries of representative monolayer Nb$_{2n+1}$Si$_n$Te$_{4n+2}$ materials. }}
\label{tab:1}
\begin{ruledtabular}
\begin{tabular}{cccccc}
$n$ & $a~(\rm\AA)$ & $b~(\rm\AA)$ & Thickness~$(\rm\AA)$ & Space group & Point Group \\ \midrule
 $1$ & 6.408 & 11.633 & 3.649 & $Pmc2_1$ & $C_{2v}$  \\
 $2$ & 6.405 & 19.590 & 3.770 & $Pmc2_1$ & $C_{2v}$  \\
 $3$ & 6.404 & 27.552 & 3.651 & $Pmc2_1$ & $C_{2v}$  \\
 $\infty$ & 6.401 & 7.962 & 3.783 & $Pbam$ & $D_{2h}$  \\
\end{tabular}
\end{ruledtabular}
\end{table}

\begin{figure}
\begin{centering}
\includegraphics[width=8.6cm]{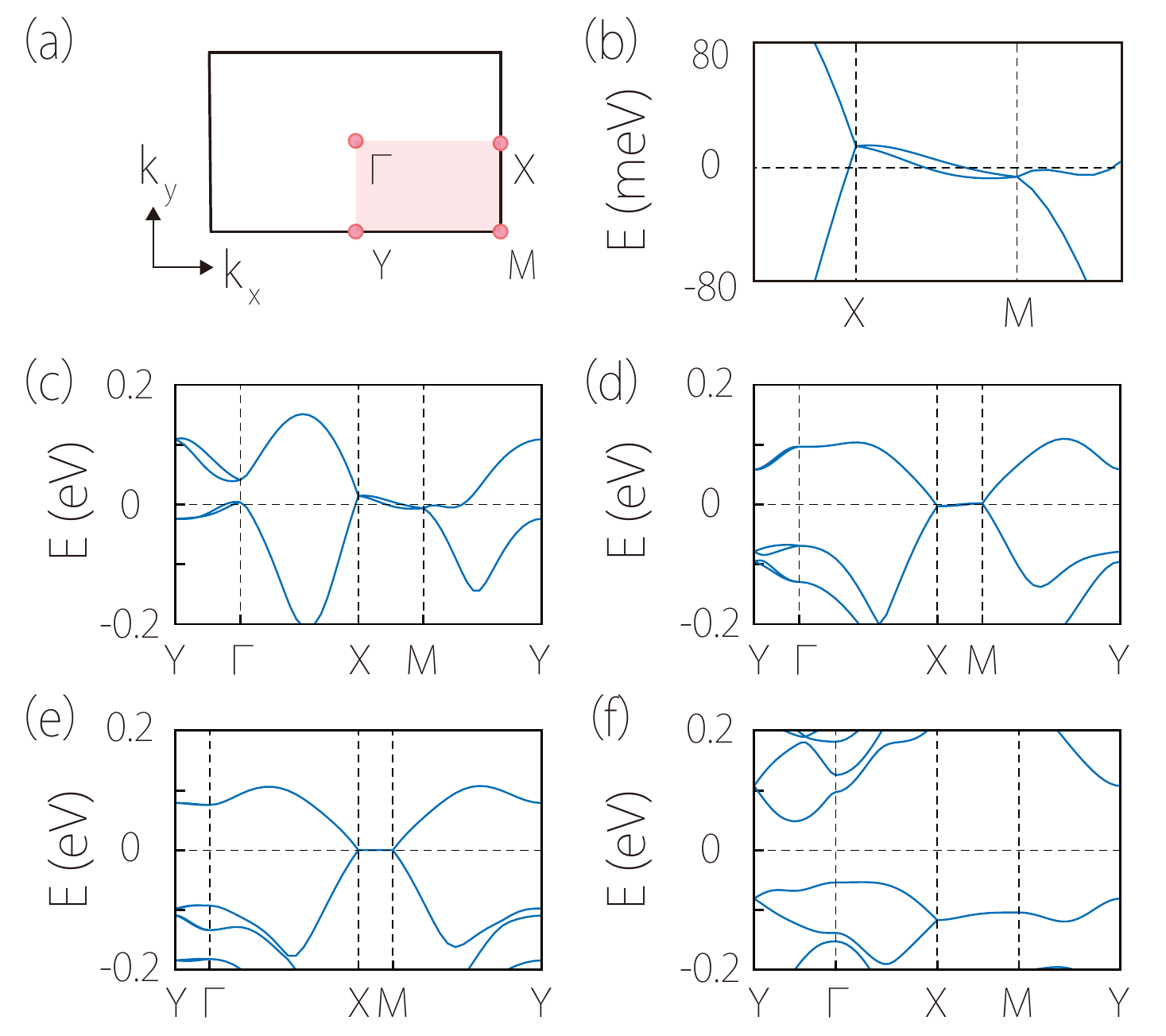}
\par\end{centering}
\caption{\label{Fig_2}{(a) Brillouin zone for monolayer Nb$_{2n+1}$Si$_n$Te$_{4n+2}$. (b-f) Band structures for monolayer Nb$_{2n+1}$Si$_n$Te$_{4n+2}$: (b, c) $n=1$, where (b) is an enlarged figure around the path $X$-$M$ in (c); (d) $n=2$; (e) $n=3$; and (f) $n=\infty$. }}
\end{figure}

In Fig.~\ref{Fig_2}, we plot the calculated electronic band structures for the four representative members in Table~\ref{tab:1}. One can see that the band structures for $n=1,2,3$ show similar features. Previous works have shown that in the absence of SOC, these materials are nodal-line semimetals~\citep{Li2018_NbSiTe,Zhu2020_NbSiTe}. The nodal line on the $X$-$M$ path around Fermi level is enforced by the nonsymmorphic $\mathcal{T}\tilde{M}_y$ symmetry. The detailed analysis was given in our previous works~\citep{cao2022plasmons}, so we will not repeat it here.
It should be noted that in Fig.~\ref{Fig_2}, the band structures include the SOC effects. Under SOC, the $\mathcal{T}\tilde{M}_y$ symmetry protection is no longer exact, so the original nodal line degeneracy will be lifted. In the enlarged view in Fig.~\ref{Fig_2}(b), one can clearly see the splitting of the nodal line. Nevertheless, there is still a degenerate nodal point at $X$ (and also at $M$). This point is a fourfold degenerate Dirac point enforced by nonsymmorphic symmetries of the system. Its formation mechanism has been discussed in Ref.~\citep{Li2018_NbSiTe}. The SOC induced change to the band structure is weak, so for many properties, SOC may just be neglected. However, band geometric properties like Berry curvature and BCD are very sensitive to small-gap regions in band structures, such as those due to SOC splitting. Therefore, to study BCD and its nonlinear Hall effect, we have to include SOC in the calculation.

 The low-energy states around Fermi level are mostly distributed on the $c$ chains. Previous scanning tunneling spectroscopy (STS) experiments also verified this feature~ \citep{Zhu2020_NbSiTe,Wang2021_NbSiTe}. With increasing $n$, the distance between two $c$ chains will increase and hence the coupling between them will decrease. As a result, the band dispersion will become flatter along the $y$ direction, as can be seen in Figs.~\ref{Fig_2}(c)-\ref{Fig_2}(e) along the $\Gamma$-$Y$ and $X$-$M$ paths.

For Nb$_2$SiTe$_4$ with $n=\infty$, Fig.~\ref{Fig_2}(f) shows that it is a narrow-gap semiconductor. {The band gap is $\sim 0.51$ eV, which is slightly larger than the band gap of layered Nb$_2$SiTe$_4$ ($\sim 0.39$ eV)~\citep{zhao2019nb2site4}.} This different character can now be understood from the discussion above. One can view the $c$ chains as metals, whereas the $(ab)$ chains are insulating. Since Nb$_2$SiTe$_4$ is entirely made of $(ab)$ chains, its spectrum would naturally be gapped.

The features discussed above, particularly the evolution of band structure with $n$, will have important implications on
BCD and nonlinear Hall response in these materials.

\section{Berry curvature dipole}
Berry curvature is an intrinsic band geometric quantity. It plays an important role in many physical properties, especially anomalous transport properties~\citep{xiao2010rmp}. In nonmagnetic materials, nonzero Berry curvature requires the breaking of inversion symmetry. This condition is fulfilled in monolayer Nb$_{2n+1}$Si$_n$Te$_{4n+2}$ with finite $n$. For Nb$_2$SiTe$_4$ with $n=\infty$, inversion symmetry is respected and hence Berry curvature vanishes identically.

For a 2D system, Berry curvature only has a single component, which can be expressed as (we set $e=\hbar=1$ in the formulas)
\begin{equation}
\label{eq1}
  \Omega_z(n\bm k)=-2\,\text{Im}\sum_{n'\neq n}\frac{\langle u_{n\bm k}|v_x|u_{n'\bm k}\rangle\langle u_{n'\bm k}|v_y|u_{n\bm k}\rangle}{(\varepsilon_{n\bm k}-\varepsilon_{n'\bm k})^2},
\end{equation}
for a state $|u_{n\bm k}\rangle$, where $v_x$ and $v_y$ are the velocity operators, and $\varepsilon_{n\bm k}$ is the energy of $|u_{n\bm k}\rangle$.

\begin{figure}
\begin{centering}
\includegraphics[width=8.6cm]{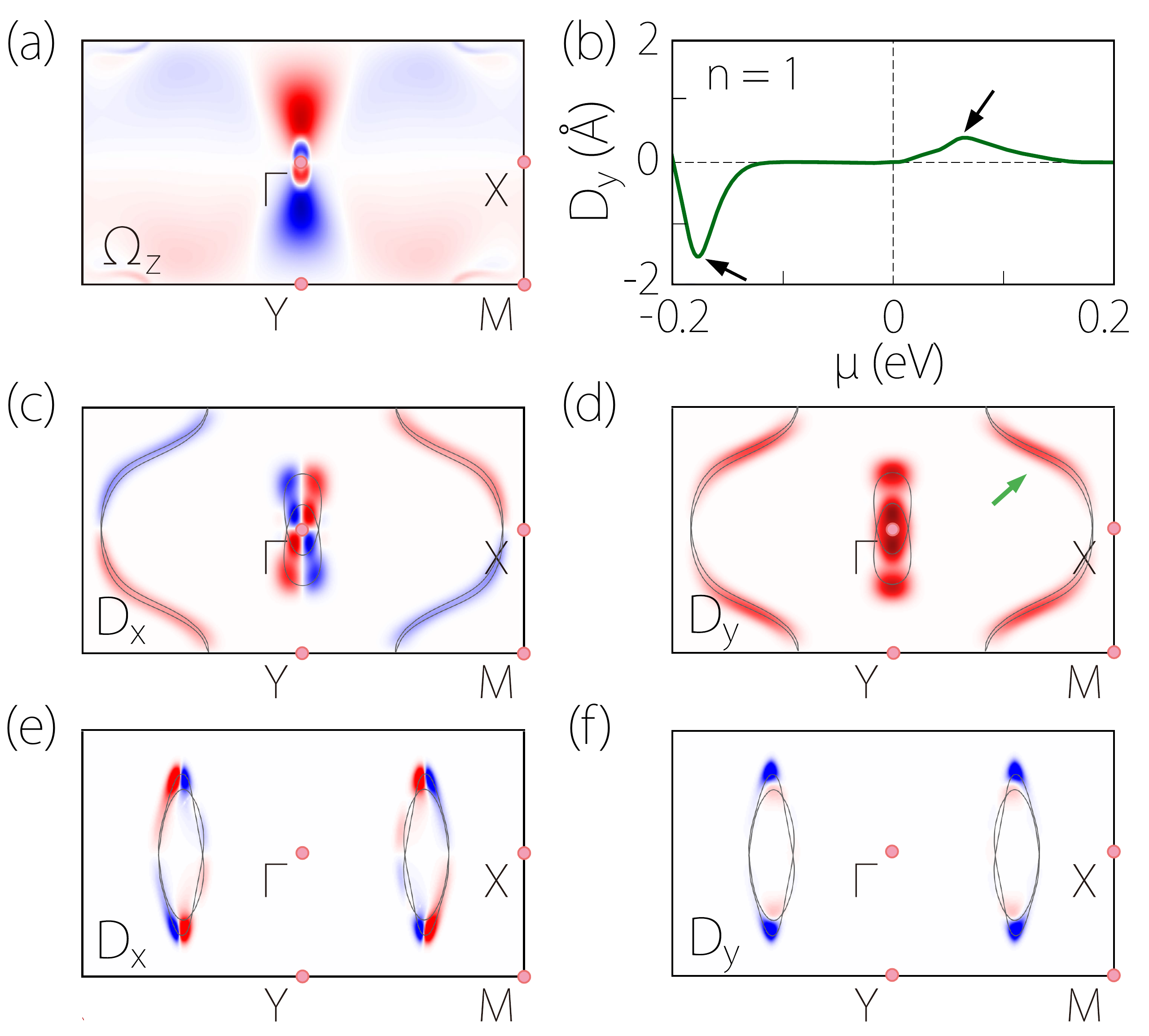}
\par\end{centering}
\caption{\label{Fig_3}{Berry curvature and its dipole (BCD) in $\rm Nb_3SiTe_6$. (a) Distribution of Berry curvature for the occupied states. (b) The BCD $\mathcal{D}_y$ versus chemical potential $\mu$. (c-f) The $k$-resolved BCD as defined in Eq.~(\ref{eq4}). Here, (c, d) are plotted for $\mu=0.064$~eV (the upper peak in (b)), and (e, f) are for $\mu=-0.180$~eV (the lower peak in (b)). The Fermi contours at these energies are indicated by the black curves. }}
\end{figure}

Consider Nb$_3$SiTe$_6$ ($n=1$). In Fig.~\ref{Fig_3}(a), we plot the distribution of its Berry curvature in BZ for occupied states, i.e., the quantity
\begin{equation}
  \Omega(\bm k)=\sum_n f_0\Omega_z(n\bm k),
\end{equation}
where $f_0$ is the Fermi distribution function. One observes that {the Berry curvature is odd in $k_y$ and even in $k_x$, as required by $\mathcal{T}$ and $\tilde{M}_y$, and its value is quite pronounced along the
$\Gamma$-$Y$ path.}

BCD is the first moment of Berry curvature in BZ. It is a pseudovector in 2D, defined as~\citep{sodemann2015quantum}
\begin{equation}\begin{split}
  {\mathcal{D}}_a=&\sum_n \int_\text{BZ}\frac{d^2k}{(2\pi)^2}f_0\partial_a\Omega_z(n\bm k)\\
  =&-\sum_n \int_\text{BZ}\frac{d^2k}{(2\pi)^2}f_0'v_a(n\bm k)\Omega_z(n\bm k),
  \end{split}
\end{equation}
where $a\in\{x,y\}$, $\partial_a\equiv\partial_{k_a}$, and in the second line, we write it as a Fermi surface integral.

For finite $n$, Nb$_{2n+1}$Si$_n$Te$_{4n+2}$ only has a single mirror line along $x$, which allows a nonzero BCD. Since $\bm{\mathcal{D}}$ is a pseudovector, it must be along the $y$ direction, i.e., $\bm{\mathcal{D}}=\mathcal{D}_y\hat{y}$. In Fig.~\ref{Fig_3}(b), we plot the calculated $\mathcal{D}_y$ versus the chemical potential $\mu$ for $n=1$. One observes two peaks in the figure: one is at 0.064 eV with a value of 0.399 $\rm\AA$, and the other is at $-0.180$ eV with a value of $-1.540$ $\rm\AA$. The two peaks are of opposite signs. We note that the magnitude of -1.540 $\rm\AA$ is quite large. This is comparable or larger than those found in {monolayer $T_d$-WTe$_2$ ($0.1\sim0.7~\rm\AA$)~\citep{you2018berry}, strained NbS$_2$ (0.2~$\rm\AA$)~\citep{xiao2020two}  and strained WSe$_2$ (0.02~$\rm\AA$)}~\citep{you2018berry}.

To understand the origin of the large BCD in monolayer Nb$_3$SiTe$_6$, in Figs.~\ref{Fig_3}(c)-\ref{Fig_3}(f), we plot the $k$-resolved BCD on Fermi surface, namely the quantity
\begin{equation}
\label{eq4}
  \mathcal{D}_a(\bm k)=-\sum_n f_0'v_a(n\bm k)\Omega_z(n\bm k),
\end{equation}
for $\mu=$ 0.064 eV (upper peak) and $-0.180$ eV (lower peak). First of all, one observes that $\mathcal{D}_x(\bm k)$ is an odd function in $k_y$ whereas $\mathcal{D}_y(\bm k)$ is an even function, as required by the $\tilde{M}_y$ symmetry. Hence, after integral over BZ, BCD only has the $y$ component left.
From  Figs.~\ref{Fig_3}(c)-\ref{Fig_3}(f), one can see that the nodal line region along $X$-$M$ does not make a sizable contribution to BCD. For the upper peak [Fig.~\ref{Fig_3}(d)], large contribution to $\mathcal{D}_y$ is from the $\Gamma$-$Y$ path, which corresponds to the SOC splitting gap indicated in Fig.~\ref{Fig_2}(c). The spin splitting gap on the outer Fermi surface [marked by the green arrow in Fig.~\ref{Fig_3}(d)] also gives a non-negligible contribution. As for the lower peak, Figs.~\ref{Fig_3}(e)-\ref{Fig_3}(f) show that the Fermi surface has two separate pieces. By examining the band structure around the hot spots in Fig.~\ref{Fig_3}(f), we find that the large negative contribution is also from SOC splitting of the band structure.

\begin{figure}
\begin{centering}
\includegraphics[width=8.6cm]{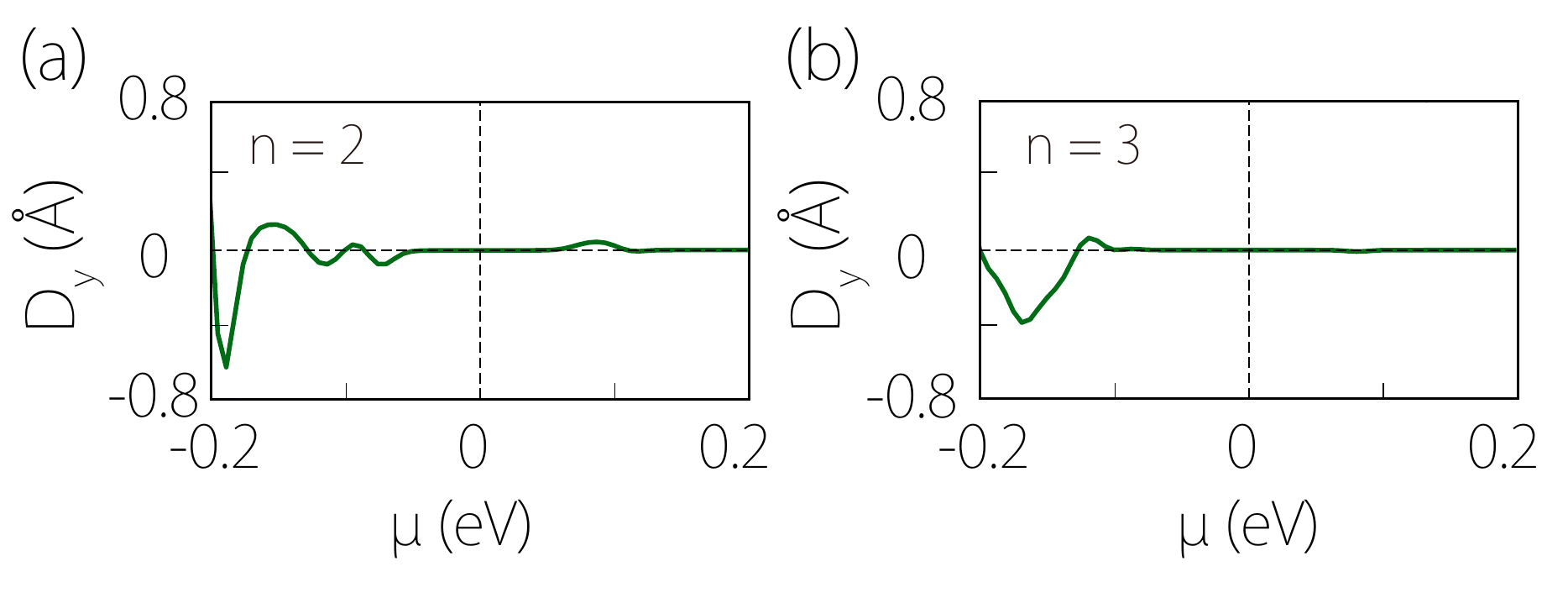}
\par\end{centering}
\caption{\label{Fig_4}{The BCD $\mathcal{D}_y$ versus chemical potential $\mu$ for (a) $n=2$, and  (b) $n=3$ cases.}}
\end{figure}

Next, we consider the cases with $n=2$ and 3. From the results in Fig.~\ref{Fig_4}, one can see that the magnitude of BCD decreases with increasing $n$. For $n=3$, the BCD value above $\mu=0$ (which is also the energy of nodal line) is already negligibly small.
As for the lower peak, the value is about 0.663 $\rm\AA$ for $n=2$ and 0.396 $\rm\AA$ for $n=3$.

\begin{figure}
\begin{centering}
\includegraphics[width=8.6cm]{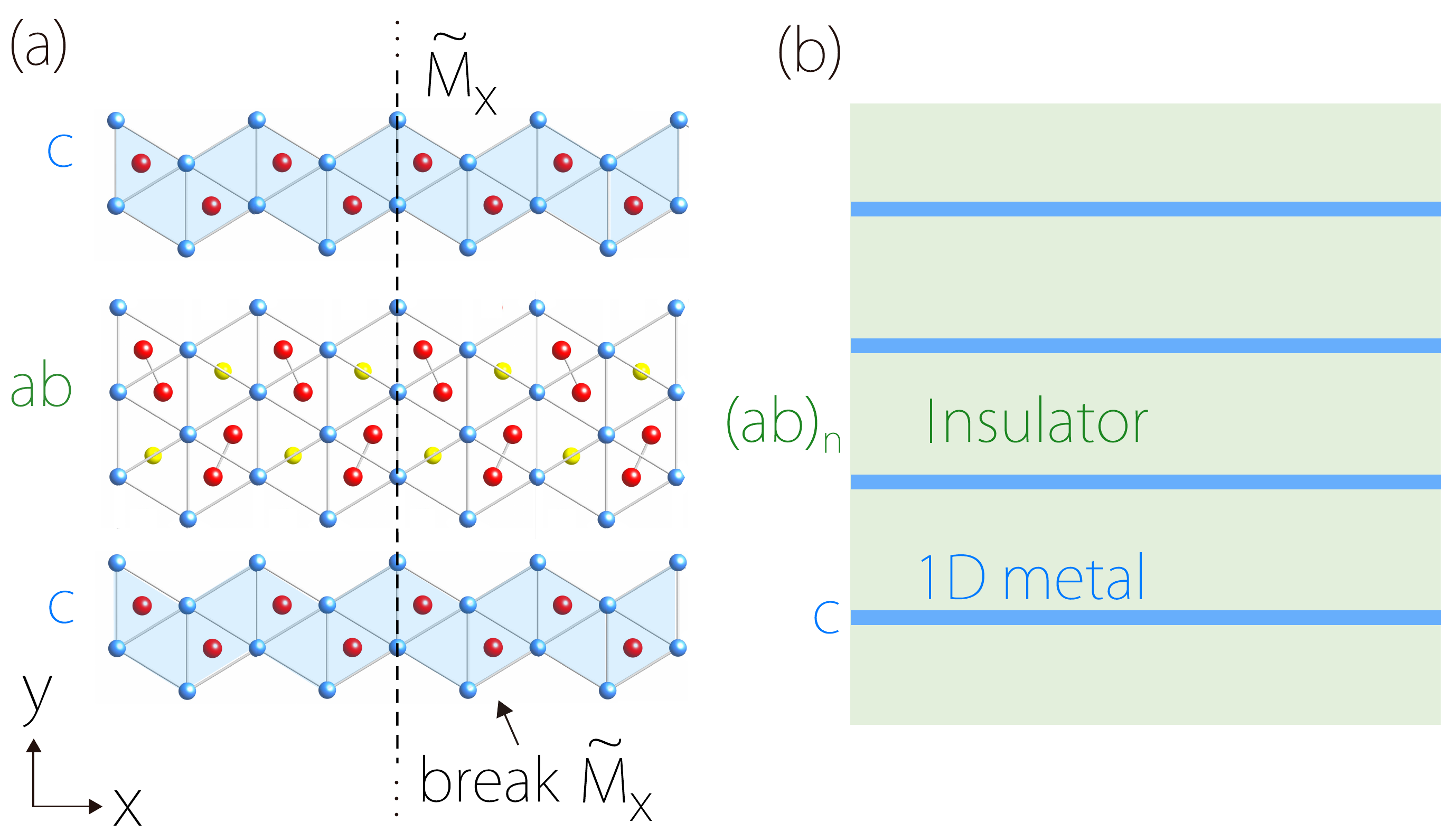}
\par\end{centering}
\caption{\label{Fig_5}{(a) $(ab)$ chains preserve the $\tilde{M}_x$ symmetry, whereas $c$ chains break it. Hence, the density of $c$ chains represents the extend of $\tilde{M}_x$ symmetry breaking.
(b) The Nb$_{2n+1}$Si$_n$Te$_{4n+2}$ system may be schematically viewed as 1D metallic chains ($c$ chains) embedded in a 2D insulator matrix (made of $(ab)$ chains).}}
\end{figure}

This trend of decreasing BCD with increasing $n$ in monolayer Nb$_{2n+1}$Si$_n$Te$_{4n+2}$ can be understood from two perspectives. First, in terms of symmetry, Nb$_{2n+1}$Si$_n$Te$_{4n+2}$ with finite $n$ supports BCD because of its low symmetry. The presence of $c$ chains is crucial because they break the $\tilde{M}_x$ symmetry of $(ab)$ chains (Fig.~\ref{Fig_5}). Without $c$ chains, $\tilde{M}_x$ becomes an exact symmetry and it suppresses BCD (given the other mirrors in the system) as in the $n=\infty$ limit. Hence, the density of $c$ chains in the system can be viewed as a measure of the extent of symmetry breaking. It is strongest in $n=1$ case, and gradually decreases as $n$ increases, determining the trend in BCD.

Meanwhile, the trend is also connected with the dimensional crossover in this system~\citep{Zhang2022_NbSiTe}. As discussed, the low-energy states are mostly distributed on the $c$ chains. One may view the $c$ chains as metallic 1D subsystems put in an insulating
matrix formed by the $(ab)$ chains. For small $n$, the system retains a 2D character, because the $c$ chains are not far from each other and the inter-chain coupling is sizable. However, with increasing $n$, the inter-chain coupling will decrease, and the system approaches the quasi-1D character. Berry curvature is a differential 2-form, which vanishes in the 1D limit [as can also be seen from Eq.~(\ref{eq1})]. Thus, BCD must decrease and approach zero during this dimensional crossover.

It must be emphasized that the dimensional crossover here is referring to the low-energy electronic states. Structurally, Nb$_{2n+1}$Si$_n$Te$_{4n+2}$ materials always maintain a 2D material character: the lattices are strongly bonded in both $x$ and $y$ directions. Thus, the crossover is a hidden feature that occurs only for the electronic sector. This is a very interesting piece of physics for 2D Nb$_{2n+1}$Si$_n$Te$_{4n+2}$ materials. Now, we revealed its manifestation in BCD, which can be detected via nonlinear Hall measurement.

\begin{figure} % [htbp]
\begin{centering}
\includegraphics[width=8.6cm]{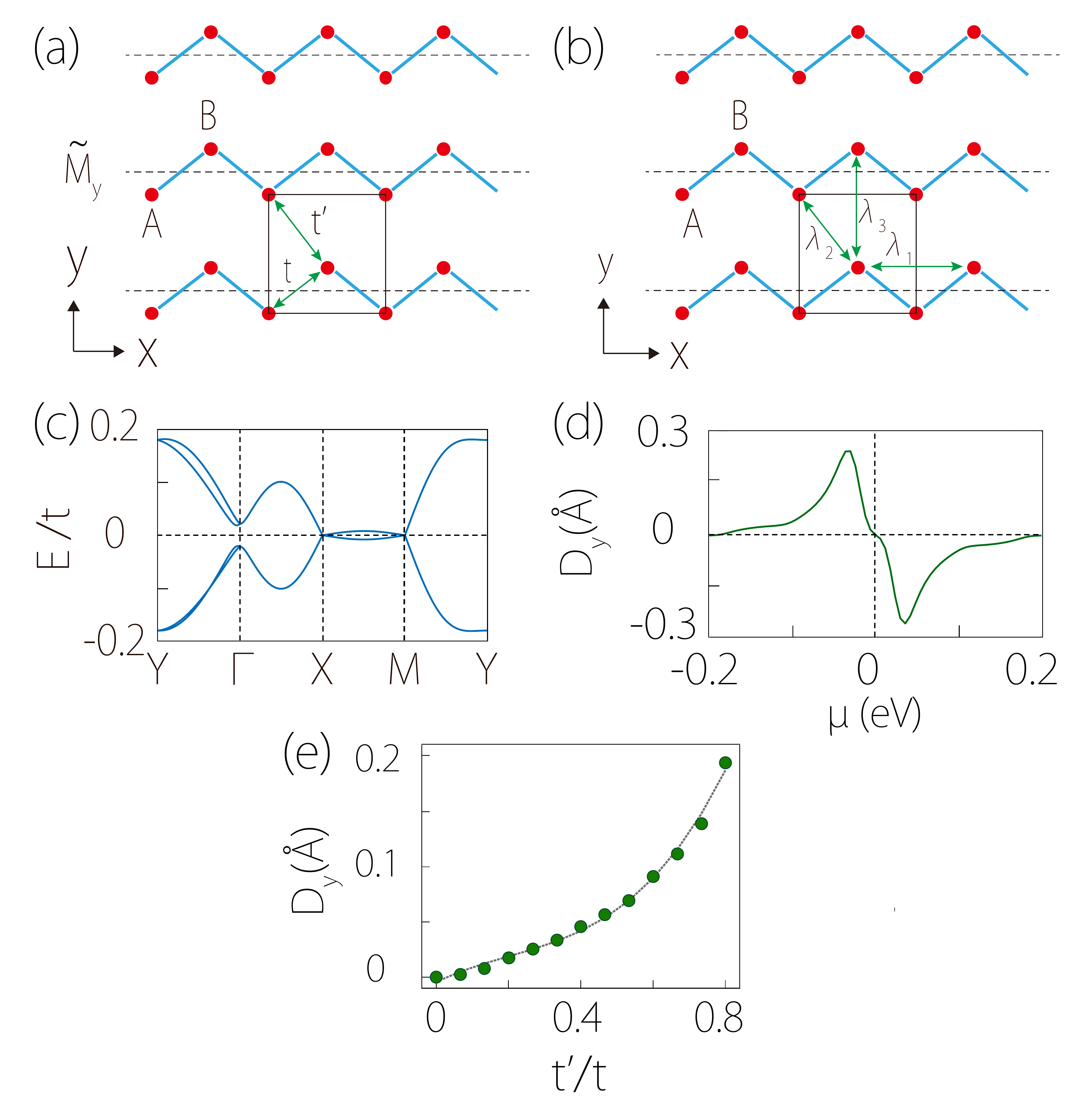}
\par\end{centering}
\caption{\label{Fig_6}{(a) Schematic figure showing the tight-binding model. The model consists of zigzag chains. A primitive cell contains two sites $A$ and $B$. $t$ and $t'$ are the amplitudes for intrachain and interchain hoppings. (b) illustrates the three hopping processes corresponding to the SOC terms in Eq.~(\ref{SOC}). (c) Band structure of the tight-binding model. (d) Corresponding BCD $\mathcal{D}_y$ versus chemical potential. (e) Variation of BCD peak value as a function of the interchain coupling. The solid curve is a guide to the eye. In (c, d), we set $t=0.2$~$\rm eV$, $t'=0.16$~$\rm eV$, $\lambda_1=1$, and $\lambda_2=\lambda_3=0.1$. The same values of $t$ and $\lambda$'s are taken in (e).}}
\end{figure}

\section{A Model study}
To understand the features in band structure and in BCD, we construct an effective lattice model to describe the
low-energy bands in monolayer Nb$_{2n+1}$Si$_n$Te$_{4n+2}$ with finite $n$. The model may also serve as a good starting point for other theoretical studies.

In Refs.~\citep{Zhang2022_NbSiTe,cao2022plasmons}, we have proposed a 2D Dirac Su-Schrieffer-Heeger (SSH) model, which is spinless (i.e., without SOC) and captures the nonsymmorphic nodal line feature in monolayer
Nb$_{2n+1}$Si$_n$Te$_{4n+2}$. However, to study BCD, as we noted, the consideration of SOC is necessary. Therefore, we need to extend the previous spinless Dirac SSH model to include SOC effects.

The Dirac SSH model is defined on a rectangular lattice, as shown in Fig.~\ref{Fig_6}.  It consists of an array of zigzag chains running in the $x$ direction. Physically, each chain corresponds to a $c$ chain in Nb$_{2n+1}$Si$_n$Te$_{4n+2}$. In a unit cell, there are two sites $A$ and $B$. Assigning one orbital at each site and considering the nearest intra-chain and inter-chain hoppings, one obtains the following model constrained by {$\mathcal{T}$, $\tilde{M}_y$, and $M_z$ symmetries:
\begin{eqnarray}
\mathcal{H}_0 & = & t\left[\begin{array}{cc}
0 & 1+e^{-ik_{x}}\\
1+e^{ik_{x}} & 0
\end{array}\right]\sigma_{0} \nonumber\\
 &  & + t^{\prime}\left[\begin{array}{cc}
0 & e^{-ik_{y}}\left(1+e^{-ik_{x}}\right)\\
e^{ik_{y}}\left(1+e^{ik_{x}}\right) & 0
\end{array}\right]\sigma_{0},
\end{eqnarray}
where the momenta are measured in unit of the inverses of lattice constants, and the Pauli matrices $\sigma$ denote the spin degree of freedom. }

Next, we add SOC to the model. The above mentioned symmetries resulted in the following SOC terms up to second neighbor hopping processes:
{\begin{equation}\begin{split}\label{SOC}
\mathcal{H}_\text{SOC} & =  t\left[\begin{array}{cc}
2\lambda_{1}\sin k_{x} & 0\\
0 & -2\lambda_{1}\sin k_{x}
\end{array}\right]\sigma_{z} \\
    +& t^{\prime}\left[\begin{array}{cc}
2\lambda_{3}\sin k_{y} & i\lambda_{2}e^{ik_{y}}\left(1+e^{-ik_{x}}\right)\\
-i\lambda_{2}e^{-ik_{y}}\left(1+e^{ik_{x}}\right) & 2\lambda_{3}\sin k_{y}
\end{array}\right]\sigma_{z}. 
\end{split}
\end{equation}
}
Here, the first term is from intrachain hopping process, whereas the second term is from interchain process, as indicated in Fig.~\ref{Fig_6}(b).  

Therefore, our spin-orbit-coupled Dirac SSH model is obtained as
\begin{equation}
  \mathcal{H}=\mathcal{H}_0+\mathcal{H}_\text{SOC}.
\end{equation}
In Fig.~\ref{Fig_6}(c), we plot a typical band structure of this model. Namely, there is an approximate nodal line on the $X$-$M$ path (split by SOC); the
SOC splitting is observed on $X$-$M$ and $Y$-$\Gamma$ paths, but not on the $\Gamma$-$X$ and $M$-$Y$ paths. {The double degeneracy on $X$-$M$ and $Y$-$\Gamma$ is due to the anti-commutation between $\tilde{M}_y$ and $M_z$ on these two paths. }
One can see that it indeed captures the main features of
DFT band structures in Fig.~\ref{Fig_2}(c). In Fig.~\ref{Fig_6}(c), we plot the BCD calculated for this model. The two BCD peaks in Fig.~\ref{Fig_3}(b) are reproduced in this simple model. One peak is above the nodal-line energy and the other one is below, and they have opposite signs.
%In Fig., we also plot the $k$-resolved BCD $\mathcal{D}_y$. One observes that ...
Finally, we plot the BCD peak magnitude as a function of interchain coupling $t'$. One can see that {the value monotonically increases with the interchain coupling.} Since $t'$ decreases with $n$ in monolayer
Nb$_{2n+1}$Si$_n$Te$_{4n+2}$, the behavior in Fig.~\ref{Fig_6}(d) agrees with our result from DFT calculations.

\section{Nonlinear Hall effect}

\begin{figure}
\begin{centering}
\includegraphics[width=8.6cm]{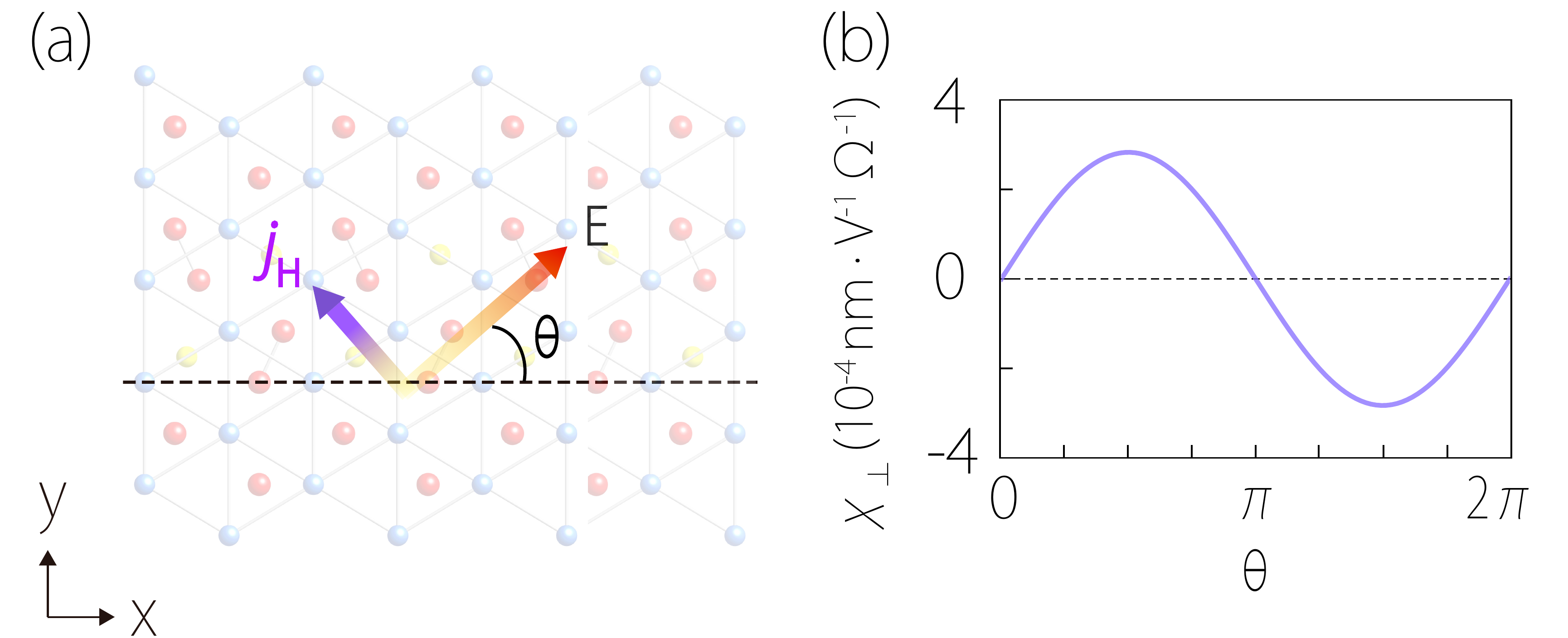}
\par\end{centering}
\caption{\label{Fig_7}{(a) Illustration of the nonlinear Hall current induced by $E$ field in Nb$_{3}$SiTe$_{6}$. The in-plane $E$ field makes an angle $\theta$ from the mirror line. The induced Hall current is perpendicular to the $E$ field, as indicated by the green arrow.   (b) Nonlinear Hall conductivity $\chi_\text{H}$ versus the angle $\theta$.}}
\end{figure}

It was shown that BCD leads to a second-order nonlinear Hall current. For a 2D system, the current can be expressed as
\begin{equation}
  \bm j_\text{H}=-\frac{1}{2}\tau \hat{z}\times\bm E(\bm{\mathcal{D}}\cdot\bm E),
\end{equation}
where $\bm E$ is the applied in-plane $E$ field, and $\tau$ is the relaxation time. Consider monolayer Nb$_{2n+1}$Si$_n$Te$_{4n+2}$ with the coordinate setup in Fig.~\ref{Fig_7}(a). Assuming applied $E$ field is in the direction specified by the polar angle $\theta$ (with respect to the mirror line), i.e.,
$(E_x,E_y)=E(\cos\theta,\sin\theta)$, then the Hall current will be in the direction of $(j_x,j_y)=j_\text{H}(-\sin\theta,\cos\theta)$, with the Hall current magnitude
\begin{equation}
  j_\text{H}=\chi_\text{H}(\theta) E^2,
\end{equation}
and the nonlinear Hall conductivity
\begin{equation}
  \chi_\text{H}(\theta)=-\frac{1}{2}\tau\mathcal{D}_y\sin\theta.
\end{equation}

Experimentally, a 2D material sample can be etched into a disk shape and attached with multiple pairs of leads~\citep{kang2019nonlinear,Lai2021_3th}, such that the $\sin\theta$ angular dependence in the nonlinear Hall response can be verified in experiment. To measure the second-order nonlinear response, one typically modulates the driving source with a low frequency and detects the signal at doubled frequency using the lock-in technique~\citep{ma2019observation,kang2019nonlinear}. The Fermi level of 2D materials can be readily tuned by using electric gating technique.
Here, consider monolayer Nb$_3$SiTe$_6$ (i.e., $n=1$). With our calculated {$\mathcal{D}_y\sim 1.54~\rm \AA$} at the lower peak, assuming {$\tau= 10~\rm ps$} which is typical for 2D materials, the magnitude of $\chi_\text{H}$ can reach {$2.9\times{10}^{-4}$~nm$\cdot$S/V} and its angular dependence is shown in Fig.~\ref{Fig_7}(b). Under a driving field of {$E\sim {10}^{4}$~V/m}, the resulting nonlinear Hall current density can reach {$\sim 0.6$~$\mu$A/cm }. For $n=2$ ($3$), the signal is expected to be smaller by a factor $\sim 2$ ($\sim 4$), which is still detectable in experiment.

\section{Conclusion}
We have revealed monolayer Nb$_{2n+1}$Si$_n$Te$_{4n+2}$ materials as a suitable platform for studying BCD and nonlinear Hall effect. These materials have the adequate symmetry to support the effect without extra strain, enjoy stability at ambient conditions, and exhibit composition tunability. We show that BCD is most pronounced for the $n=1$ case, where its magnitude can reach  {$1.54~\rm \AA$}. The BCD value decreases with increasing $n$. This can be understood from degree of symmetry breaking and also from a dimensional crossover. It is interesting that this crossover occurs only for the low-energy electronic states, whereas
structurally, the system is always strongly bonded in 2D. The evolution of BCD with $n$ can be regarded as a manifestation of this hidden transition. We construct the spin-orbit-coupled Dirac SSH model, which captures the main features of DFT results. The nonlinear Hall conductivity and its angular dependence are analyzed. Our work uncovers interesting geometric quantities and  nonlinear physics in Nb$_{2n+1}$Si$_n$Te$_{4n+2}$ family materials. It provides useful guidance for subsequent experiments on these systems.

\begin{acknowledgments}
The authors thank D. L. Deng for helpful discussions. This work is supported by Singapore MOE AcRF Tier 2 (T2EP50220-0026), National Natural Science Foundation of China (Grant Nos.~52271136, 11704304, 12204378), and Natural Science Foundation of Shaanxi Province (Grant Nos.~2019TD-020, 2019JLM-30, 2017JZ015, 2018JQ1028). The computing for this work was performed at the High Performance Computing Center at Xi’an Jiaotong University. 
\end{acknowledgments}

\bibliographystyle{apsrev4-2}
\bibliography{ref}

\end{document}